\documentclass [preprint,aps,showpacs]{revtex4}
\usepackage[final]{graphics}
\usepackage{amssymb}
\usepackage{amsfonts}
\usepackage{epsfig}
\usepackage{graphicx}

\begin{document}
\title{Moment ratios for the pair contact process with diffusion}
\author{Marcelo M. de Oliveira\footnote{mancebo@fisica.ufmg.br}
and Ronald Dickman\footnote{dickman@fisica.ufmg.br}}
\address{Departamento de F\'{\i}sica, Instituto de Ci\^encias Exatas,\\
Universidade Federal de Minas Gerais \\
C. P. 702, 30123-970, Belo Horizonte, Minas Gerais - Brazil }

\date{\today}

\begin{abstract}
We study the continuous absorbing-state phase transition in the one-dimensional
pair contact process with diffusion (PCPD).  In previous studies [Dickman and
de Menezes, Phys. Rev. E, {\bf 66} 045101(R) (2002)], the critical point moment ratios of
the order parameter showed anomalous behavior, growing with system size rather
than taking universal values, as expected. Using the quasistationary simulation
method we determine the moments of the order parameter up to fourth order at
the critical point, in systems of up to 40960 sites. Due to strong finite-size
effects, the ratios converge only for large system sizes. Moment ratios and
associated order-parameter histograms are compared with those of directed
percolation.  We also report an improved estimate [$p_c =0.077092(1)$] for the
nondiffusive pair contact process.
\end{abstract}


\maketitle

\section{Introduction}

The pair contact process (PCP) \cite {jensen93} is a nonequilibrium stochastic
model that exhibits a phase transition to an absorbing state
\cite{marro,hinrichsen,lubeck04}. Several studies have established that the PCP
belongs to the robust directed percolation (DP) universality class
\cite{odor04}.  In contrast with the contact process, which has the vacuum as
its unique absorbing state, the PCP exhibits an infinite number of absorbing
configurations, since both creation and annihilation require a nearest-neighbor
pair of particles. Allowing particles in the PCP to hop on the lattice, one
obtains the so-called pair contact process with diffusion (PCPD).  Here there
are only two absorbing states: the vacuum, and the subspace with only a single
particle.

While a version of the pair contact process with diffusion was proposed by
Grassberger in 1982 \cite{grassberger82}, current interest in the problem
follows its rediscovery by Howard and Ta\"uber \cite{how-tau97} who suggested
that its Langevin description involves complex noise. The first numerical
studies of the PCPD \cite{carlon} suggested that critical the PCPD would fall
in the ``parity conserving" (PC) class, but increasing computational effort
revealed that this was not the case, and that the critical behavior of the
model was masked by huge corrections\cite{hinri01a,odor00}. Since then, diverse
scenarios, some of them contradictory, have been proposed in order to clarify
this question. At present there are two principal schools of thought: in one,
the PCPD belongs to a novel universality class distinct from DP, with a unique
set of critical exponents, or possibly continuously varying exponents due to a
marginal perturbation \cite{noh-park04a,dic-arg,parkpark1}. The opposing school
holds that the PCPD should be attracted to a DP fixed-point after a huge
crossover time \cite{hinri03,hinri06}. A recent review on these and other
scenarios can be found in \cite{henkel}. In this work we study the PCPD via
quasistationary (QS) simulations \cite{qssim}, focussing on the order parameter
and its moments.

The balance of this paper is organized as follows: In the next section we
review the definition of the model and detail our simulation method. In Sec.
III we present our results, and Section IV is devoted to discussion and
conclusions.

\section{Model and Simulation Method}

The PCP is defined on a lattice, with each site either occupied or vacant.
All changes of configuration involve a pair of particles occupying nearest-neighbor
sites, called a {\it pair} in what follows.
A pair annihilates itself at a
rate of $p$, and with rate $1-p$ creates a new particle at a randomly
chosen site neighboring the pair, if this site is vacant. In the PCPD, in
addition to the creation and annihilation processes already mentioned, each
particle attempts to hop, at rate $D$, to a randomly chosen nearest-neighbor
(NN) site; the move is accepted if the target site is vacant. The PCPD exhibits
a continuous phase transition to the absorbing state, at a critical
annihilation rate $p_c(D)$. Several variants of the model, differing in how
each process (creation, annihilation or diffusion) is selected, have been
studied in the literature \cite{henkel}.  Here we use the implementation of
Ref. \cite{dic-arg}. The model is defined on a ring of $L$ sites.
We maintain a list of pairs to improve efficiency. At each step
of the evolution we first choose between diffusion (with probability $D$) and
reaction (with probability $1-D$). In a reaction step, we select a
pair at random, and choose between creation and annihilation
with probabilities $p$ and $1-p$, respectively.
If we choose diffusion, a particle is selected at random; it attempts to hop to
one of its neighbor sites (if the latter is vacant). The time increment
associated with each step is $\Delta t = 1/(N_{pair} + DN_{part})$, where
$N_{pair}$ and $N_{part}$ are the number of pairs and particles in the lattice
at time $t$. Thus each lattice site is effectively visited once (on average) per
time unit. The most obvious definition of the order parameter in the PCPD is the {\it pair
density} $\rho_2$, the number of pairs per site.  Since creation (and destruction)
of particles requires pairs, one might expect the particle density $\rho_1$ to
scale in a similar manner.

In the studies reported here we sample the {\it quasistationary}
(QS) distribution of the process, (that is, conditioned on
survival), which is very useful in the study of processes with an
absorbing state.
(In fact, conventional simulations of
``stationary" properties of lattice models with an absorbing state
actually study the quasistationary regime, given that the only
true stationary state for a finite system is the absorbing one).
We employ a recently devised simulation method
that yields quasistationary properties directly \cite{qssim}.
This is done by maintaining, and gradually updating, a set of
configurations visited during the evolution; when a transition to
the absorbing state is imminent the system is instead placed in
one of the saved configurations. Otherwise the evolution is
exactly that of a conventional simulation.

The above scheme was shown \cite{qssim} to yield precise results,
in accord with the exact QS distribution for the contact process
on a complete graph, and with conventional simulations of the same
model on a ring \cite{qssim}. The scheme has also been shown to
yield results that agree, to within
uncertainty, with the corresponding results of conventional
simulations for a sandpile model \cite{qssand}.
The advantage of
the method is that a realization of the process can be run to arbitrarily long
times. Thus, whereas in conventional simulations a large number of realizations
must be performed to have a decent sampling of the (quasi)stationary state,
here every realization provides useful information, once the initial transient
has relaxed.  This leads to an order of magnitude improvement in efficiency, in
the critical region.  For further details on the method see \cite{qssim}.

The PCPD dynamics is characterized by (1)
single-particle diffusion and (2) reactions involving a
pair, motivating us to ignore the ``purely diffusive" subspace. This means that
we modify slightly the dynamics of the model, {\it restricting it to the subspace with
at least one pair}, which we call the {\it reactive} subspace.
The motivations for studying the modified process are: (1) In \cite{dic-arg},
excluding the subspace without pairs yielded better behaved moment ratios; (2)
eliminating the large fraction of time expended on the nonreactive intervals
yields a major improvement in efficiency.  We may further justify our choice
by noting that any scaling properties necessarily involve reactions, that is,
the existence of pairs.

We impose this restriction as follows.  The initial configuration
(all sites occupied) has a large number of pairs.
Applying the QS simulation method, we accumulate a list of configurations
during the evolution.  Then, whenever a visit to a configuration (absorbing or not)
without pairs is imminent, the system is instead placed in a configuration
selected randomly from the list.  In other words, the QS method, previously
used to sample quasistationary properties, is used here to {\it restrict}
the dynamics to the subspace of interest. A subtle difference between the
two applications of the method is that, while in
usual QS simulations \cite{qssim} the method provides a just sampling of
properties conditioned on survival, in the present case we eliminate certain
nonabsorbing configurations (and the histories involving them)
as well.  For the reasons given above, we do not
expect this to color our results for large system sizes.

\section{Simulation results}

We performed extensive simulations of the PCPD
on rings of $L=1280,2560,...40960$ sites, using the QS method
restricted to the reactive subspace. Each
realization of the process is initialized with all sites
occupied, and runs for $10^9$ time steps. Averages are
taken in the QS regime, after discarding an initial transient
which depends on the system size.  In practice we accumulate
histograms of the time during which the system has exactly 1,
2,...n,... pairs, and similarly for particles.  The histograms are
used to evaluate moments; we denote by $m_{j;1}$ the $j$-th moment
of the particle number probability distribution; $m_{j,2}$ denotes
the corresponding moment of the pair number distribution. (Thus the
order parameter $\rho_2$ could also be denoted $m_{1;2}$.)
The QS lifetime $\tau$ is
taken as the mean time between attempts to leave the
reactive subspace.

The number of saved configurations ranges from 10000, for $L=1280$,
to $500$ for $L=40960$.  Values of $p_{rep}$ range from $10^{-4}$
to $2\times 10^{-6}$. The results of QS simulations were
found to agree with results of conventional simulations
\cite{dic-arg}, for system sizes $L = 80, 160...1280$. We study
three diffusion rates, $D=$ 0.1, 0.5 and 0.85, as in
\cite{dic-arg}. For comparison, we also study the (nondiffusive)
PCP, whose scaling properties are known to be those of directed
percolation. The QS results for the $D=0$ case confirm DP
behavior in the nondiffusive PCP, as can be verified from the
values of the exponents and moment ratios listed on Tables I and II.
Further, we improved the estimate for the
critical point of the PCP, obtaining $p_c=0.077092(1)$.
(This is consistent with the previous best estimate, $p_c = 0.077090(5)$,
of Ref. \cite{dic-jaf}.)

The first step in analyzing our results is to determine, for each $D$ value
studied, the critical annihilation probability $p(D)$. We use the following
criteria for criticality: power-law dependence of (1) $\rho_1$ and $\rho_2$ and
(2) $\tau$ on system size $L$ (i.e., the usual finite-size scaling relations
$\rho \sim L^{-\beta/\nu_\perp}$ and $\tau\sim L^z$); (3) constancy of the
moment ratio $r_{211;2} \equiv m_{2;2}/m_{1;2}^2$ with system size. The three
criteria were found to be mutually consistent within the error margins. Figure
1 shows the data for $\rho_2$ for $D=0.5$; the QS order parameter for various
$p$ values near $p_c$ is plotted versus $L$ on log scales. The data for the
four largest sizes are well fit by a straight line of slope $-\beta/\nu_\perp =
-0.385$. Plotting $L^{\beta/\nu_\perp} \rho_2$  (see inset), allows us to
eliminate as off-critical $p$ values for which the plot shows a significant
curvature, leading to the estimate $p_c= 0.120353(2)$. A similar analysis of
the particle density $\rho_1$ yields $p_c = 0.120357(5)$ while that for the
lifetime $\tau$ gives 0.120352(3), leading to the overall best estimate $p_c
0.120354(3)$ for $D=0.5$.

Analysis the moment ratio is also useful in setting limits on
$p_c$: As can be seen in Fig.2, this quantity appears to grow with
system size  for $p<p_c$, and vice-versa.
For $D=0.85$ for example, we find $p_c=0.12992(1)$ from analysis of $\rho_2$,
$p_c=0.12993(1)$ from analysis of $\tau$, and $p_c=0.129925(8)$
from analysis of the moment ratio $r_{211;2}$.
(For $D=0.5$, on the other hand, the moment ratio data do not yield an estimate
of precision comparable to that furnished by $\rho_1$, $\rho_2$ and $\tau$.)

As a check on our procedure for determining $p_c$, we also performed,
for $D=0.85$ initial decay studies
\cite{kockelkoren,parkpark06}.  In these studies the order parameter
$\rho_2$ is followed as a function of
time, starting, as in the other studies, with all sites occupied.
(In this case the system does not leave the reactive subspace on
the time scale of the simulation so the QS procedure is not
needed.) Here the expected behavior is $\rho_2 \sim t^{-\theta}$.
Using deviations from the power law to identify off-critical
values, a study of systems of $10^6$ sites, to a maximum time of 10$^8$, yields
$p_c=0.129915(15)$, fully consistent with the QS results.
Analysis of the data for $t \geq 5 \times 10^5$ furnishes $\theta=0.19(1)$, consistent
with previously reported results \cite{kockelkoren,parkpark1}.

Our present estimates for $p_c$ are slightly lower than those
reported in \cite{dic-arg}, a difference of less than $0.1\%$.
These differences highlight the strong finite-size corrections
affecting the PCPD (in \cite{dic-arg} system sizes range from 80 to
1280), and appear in other absorbing-state problems, such as the
restricted sandpile model \cite{qssand}.

Using our results for the three largest system sizes ($L=10240$,
20480 and 40960) we obtain estimates for the critical exponent
ratios $\beta/\nu_\perp$ and $z = \nu_{||}/\nu_\perp$. These are
reported in Table I. The chief contribution to the uncertainties
in the exponent ratios comes from the uncertainty in $p_c$.
We note that although the exponent $z$ appears to depend systematically
on diffusion rate $D$, our results are in fact consistent with $z=2$ in all
cases.  By contrast, the value of $\beta/\nu_\perp$ for $D=0.1$ is significantly different 
than that found for $D=0.5$ and 0.85.  The latter value ($\beta/\nu_\perp = 0.385(11)$)
does not yield a good fit to the data for $D=0.1$ even if a logarithmic correction 
term is introduced in the fitting function.

In nonequilibrium statistical physics, obtaining values for moment
ratios has proved an efficient method for identifying the
universality class \cite{dic-jaf}.  Here we analyze, in addition
to $r_{211;2}$ mentioned above, the ratios
\begin{equation}
r_{312;2} \equiv \frac{m_{3;2}} {m_{2;2} m_{1;2}}
\end{equation}
and
\begin{equation}
r_{42;2} \equiv \frac{m_{4;2}} {m_{2;2}^2}
\end{equation}
and the corresponding ratios for particles. Scaling arguments
imply that such ratios such assume universal values at the
critical point, as has been amply verified in equilibrium, and for
models with an absorbing state such as the contact process and the
PCP \cite{dic-jaf}.  Analysis of such ratios is of particular
interest in the present context, given the
perplexing result of Ref. \cite{dic-arg}, namely, apparently
{\it unlimited growth} in $r_{211;2}$ with increasing system size,
suggesting that the PCPD does not follow the usual scaling
behavior observed at absorbing-state phase transitions.
The present large-scale simulations show that the PCPD moment ratios
do in fact exhibit the expected behavior.

Figure 3 shows the moment ratios for pairs and particles in the critical PCPD,
with $D=0.5$ and the PCP.   On the basis of these data we estimate $\lim_{L \to
\infty} r_{211;2} = 1.166(8)$ for the PCPD with $D=0.5$. While the results for
$r_{211;2}$ appear to converge for the system sizes studies, the data for the
{\it particle} moment ratio $r_{211;1}$ continue to grow with system size and
cannot be extrapolated to the infinite-size limit.  (We note in passing that
the values for particles and pairs suggest that the two ratios become equal at
a system size on the order of $5 \times 10^5$, for $D=0.5$.  Since universality
implies similar scaling of particle and pair densities at the critical point,
we may take this as an order of magnitude estimate for the system size at which
corrections to scaling are finally superseded. The moment ratio data for
$D=0.85$ are consistent with this estimate, while those for $D=0.1$ suggest
convergence only at even larger sizes, on the order of $L \sim 10^7$.) Table II
lists estimates of the the limiting ($L \to \infty$) moment ratios for the PCPD
and some other known universality classes.  The ratios for $D \geq 0.5$ are
rather similar to those for directed percolation.  The cumulant ratio
$K_4/K_2^2$ (here $K_j$ denotes the $j$-th cumulant of the order parameter
probability distribution) is not in good accord with the DP value, although the
estimates for he PCPD are rather imprecise.

Another quantity that is useful for determining the universality class is the
scaled order parameter histogram, defined as $p^*(n^*) = \overline{n}p(n)$,
where $n$ denotes the number of pairs and $\overline{n}$ its mean value.  (Here
$n^* \equiv n/\overline{n}$.) Figure 4 compares the scaled histograms for the
PCPD with $D=0.1$, 0.5 and 0.85. The latter two are quite similar, while the
curve for low diffusion rate has a somewhat narrower peak.  In the inset of
Fig. 4 we compare the scaled histogram for the PCPD with $D=0.5$ to that of the
critical contact process and PCP.  While the CP and PCP have virtually
identical histograms, it is clear that even for the largest sizes studied the
PCPD histogram is very different.  (A striking difference between the PCPD on
one hand and CP/PCP on the other is the behavior near $n^* = 0$: in the latter
case the probability increases linearly with $n$, while former exhibits a
distinctly parabolic shape.  The parabolic form is not an artefact of the QS
simulation method, as it is already observed in Ref. \cite{dic-arg}, using
conventional simulation.)

In the case of $D=0.85$, we accumulated data for several values of the
annihilation rate $p$ in the vicinity of $p_c$.  This permits us to estimate
the correlation length exponent $\nu_\perp$, using the finite-size scaling
relation

\begin{equation}
m(\Delta,L)\propto \mathcal{F}_m(L^{1/\nu_\perp}\Delta),
\end{equation}
where $\Delta = p-p_c$ and $\mathcal{F}_m$ is a scaling function. This leads to
\begin{equation}
\left| \frac{\partial m}{\partial p}\right|_{p_c} \propto L^{1/\nu_\perp}.
\end{equation}

\noindent The data of Fig. 5 yield the estimate $\nu_\perp=1.09(1)$, confirming
the results of \cite{dic-arg}.

\section{Discussion}

We perform large-scale simulations of the pair contact process with diffusion,
and find that ratios of various moments of the order parameter approach
finite limiting values for large system sizes, allowing us to present the 
first numerical estimates
of such ratios for this model.   This resolves the apparent anomaly reported in
\cite{dic-arg}, of a moment ratio growing without limit.

Two basic questions prominent in recent discussions of the PCPD are: (1) Does
the model exhibit continuously variable critical exponents, or a unique set,
independent of diffusion rate $D$? (2) In the latter case, are the exponents
those of directed percolation? Despite the large computational
effort, we are unable to answer these questions definitively.  Our results
nevertheless provide some clues. With regard to the first question, the results
reported in Tables I and II show very good agreement for the two larger
diffusion rates studied ($D=0.5$ and 0.85), but those for $D=0.1$ are
quite different. Our values for the critical exponents are consistent with
earlier studies (\cite{dic-arg,noh-park04a} for low $D$ and
\cite{dic-arg,odor03} for high $D$).

For low diffusion rate ($D=0.1$) we find evidence of stronger corrections to scaling, 
and/or a slower rate of
convergence with increasing $L$, as noted above in the analysis of the moment
ratios. Our results of $z=2.08(15)$ and $\beta/\nu_\perp=0.505(10)$ are
consistent with previous results for low-diffusion rate
\cite{dic-arg,noh-park04a}. While this argues in favor of nonuniversal
behavior, we cannot rule out the possibility of a unique set of exponents, as
suggested for example by \'Odor \cite{odor03}, who asserted that, including a logarithmic
correction, the same value for $\beta/\nu_{||}$ holds for both high and low
diffusion rates.  (We note, however, that we were unable to fit all the data using the
same value for $\beta/\nu_\perp$, even including a logarithmic correction.)
In our opinion, it is still unclear if the observed difference
in the exponents and moment ratios for $D=0.1$, as compared with larger
diffusion rates, is due to corrections to scaling.  Such corrections would have
to be exceptionally large to account for the observed differences, and have to
exhibit a rather irregular behavior, given the close agreement in the results
for $D=0.5$ and 0.85.  An equally if not more natural interpretation is that
the critical exponents and moment ratios really do depend on $D$.

If we accept, provisionally, that the PCPD critical properties are
independent of $D$ (with huge corrections to scaling for small diffusion
rates), it is still not obvious that these properties are characteristic of the
DP universality class. The values for $\beta/\nu_\perp$ and $z$ are quite far
from those of DP, even for larger diffusion rates, where corrections to scaling
appear to be less severe.  It is true that our estimates for $\nu_\perp$ (for
$D=0.85$), and for the moment ratios (for both $D=0.5$ and 0.85) are close to
the DP values, but the qualitatively different form of the order parameter
histograms again prevents identification of the PCPD as belonging to
the DP universality class, even for higher diffusion rates.  Thus our results tend
to support the conclusion of Ref. \cite{parkpark06}, that PCPD scaling is 
distinct from that of DP.

Large-scale studies of time-dependent behavior \cite{kockelkoren} suggested a value
for the initial decay critical exponent $\theta$ different from that of DP.
A subsequent study \cite{hinri06} led to the suggestion
that apparent values of critical exponents vary as one increases the 
simulation time and system size, reflecting strong corrections to scaling,
and that observed exponent values be interpreted as upper limits on the true values.
The results of the quasistationary simulations are not affected by finite-time corrections,
although finite size effects are still present.

In summary, we have applied the quasistationary simulation method to the pair
contact process with diffusion, restricted to the subspace with at least one
pair. Our results indicate that the anomalous behavior of the critical 
order-parameter moment ratios does not persist for large systems.
Restricting the dynamics to the reactive sector, these ratios appear to
converge to finite values, as expected.  Taken at face value our results imply
a variation of scaling properties with diffusion rate, but the opposite
interpretation is tenable if one invokes strong corrections to scaling for
small $D$.  Regardless of whether or not the critical exponents vary with $D$,
it seems premature to conclude that the PCPD will eventually cross over to the
DP class.

\vspace{1em}

\noindent{\bf Acknowledgments}

We are grateful to G\'eza \'Odor and Haye Hinrichsen for helpful comments.
This work was supported by CNPq and FAPEMIG, Brazil.


\bibliographystyle{apsrev}

\begin{thebibliography}{100}

\bibitem{jensen93}
        I. Jensen, Phys. Rev. Lett. {\bf 70}, 1465 (1993).

\bibitem{marro}
        J. Marro and R. Dickman,
        {\it Nonequilibrium Phase Transitions in Lattice Models}
        (Cambridge University Press, Cambridge, 1999).

\bibitem{hinrichsen}
     H. Hinrichsen,
     Adv. Phys. {\bf 49} 815 (2000).

\bibitem{lubeck04}
        S. L\"ubeck,
        Int. J. Mod. Phys. B {\bf 18}, 3977 (2004).

\bibitem{odor04}
        G. \'Odor,
        Rev. Mod. Phys {\bf 76},  663 (2004).

\bibitem{grassberger82}
        P. Grassberger,
        Z. Phys. B {\bf 47}, 365 (1982).

\bibitem{how-tau97}
     M. J. Howard and U. C. T\"{a}uber,
     J. Phys. A {\bf 30}, 7721 (1997).

\bibitem{carlon}
     E. Carlon, M. Henkel and U. Schollw\"{o}ck,
     Phys. Rev. E, {\bf 63} 036101 (2001).

\bibitem{hinri01a}
     H. Hinrichsen,
     Phys. Rev. E, {\bf 63} 036102 (2001).

\bibitem{odor00}
     G. \'Odor,
     Phys. Rev. E, {\bf 62} R3027 (2000).

\bibitem{dic-arg}
     R. Dickman and M. A. F. de Menezes,
     Phys. Rev. E, {\bf 66} 045101(R) (2002).

\bibitem{noh-park04a}
     J. Noh and H. Park,
     Phys. Rev. E, {\bf 69} 016122 (2004).

\bibitem{parkpark1}
    S.-C. Park and H. Park,
    Phys. Rev. Lett. {\bf 94}, 065701 (2005).

\bibitem{hinri03}
    H. Hinrichsen, Physica A {\bf 320} 249, (2003).

\bibitem{hinri06}
    H. Hinrichsen, Physica A {\bf 361} 457, (2006).

\bibitem{henkel}
    M. Henkel and H. Hinrichsen,
    J. Phys. A: Math. Gen. {\bf 37}, R117 (2004).

\bibitem{qssim}
     M. M. de Oliveira and R. Dickman,
     Phys. Rev. E, 016129 (2005).

\bibitem{qssand}
     R. Dickman,
     Phys. Rev. E, {\bf 73} 036131 (2006).

\bibitem{dic-jaf}
     R. Dickman and J. K. Leal da Silva,
     Phys. Rev. E, {\bf 58} 4266 (1998).

\bibitem{kockelkoren}
     J. Kockelkoren and H. Chat\'e,
     Phys. Rev. Lett., {\bf 90} 125701 (2003).

\bibitem{parkpark06}
    S.-C. Park and H. Park,
    Phys. Rev. E, {\bf 73} 025105(R) (2006).

\bibitem{odor03}
        G. \'Odor,
        Phys. Rev. E, {\bf 67} 016111 (2003).

\bibitem{jensendp}
      I. Jensen,
      J. Phys. A {\bf 32}, 5233 (1999).




\end{thebibliography}

\newpage

\begin{center}
{\sf Table I. Critical exponent values for the PCP, PCPD and DP. DP values from
Ref. \cite{jensendp}.} \vspace{1em}

\begin{tabular}{c l l l}
\hline \hline
 $D$ & $p_c $ & $\beta/\nu_\perp$ &    $z$ \\
\hline
 0 (PCP)    &  0.077092(1)  & 0.2519(3)  & 1.584(7)  \\

 0.1        &  0.106405(15) & 0.505(10)  & 2.08(15)  \\

 0.5        &  0.120354(3)  & 0.385(11)  & 2.04(5)   \\

 0.85       &  0.129925(8)  & 0.386(5)   & 1.88(12)  \\
\hline
 DP  & -                    & 0.25208(5) & 1.5807(1) \\

\hline \hline
\end{tabular}
\end{center}
\vspace{3em}

\begin{center}
{\sf Table II. Critical moment ratio values for the PCP, PCPD and the DP,
parity-conserving (PC), and conserved-DP (C-DP) universality classes.}
\vspace{1em}

\begin{tabular}{c l l l l l }
\hline \hline
 D      & $r_{211;2}$ & $r_{312;2}$ & $r_{422;2}$ & $K_4/K_2^2$ & Reference \\
\hline

0 (PCP) &  1.1738(2) & 1.303(3)& 1.558(2) & -0.493(3) & \cite{dic-jaf} \\

0.1     & 1.140(15) & 1.27(2)  & 1.55(3)  & 0.1(2) & this work \\

0.5     & 1.166(8)  & 1.310(15)& 1.61(2)  & 0.0(1) & this work \\

0.85    & 1.170(6)  & 1.31(1)  & 1.61(4)  & -0.1(1) & this work \\

\hline

DP    &  1.1736(2) & 1.301(3) &  1.554(2) & -0.505(3) & \cite{dic-jaf}\\

PC    & 1.3340(4)  &          &           &           &  \cite{dic-arg} \\

C-DP  & 1.142(8)   &          &           &           & \cite{qssand} \\

\hline \hline
\end{tabular}
\end{center}

\newpage

\noindent FIGURE CAPTIONS \vspace{1em}

\noindent FIG. 1. Quasistationary order parameter versus system size for
$p=0.120345$, $p=0.120350$, $p=0.120355$ and $p=0.120360$, from top to bottom.
$D=0.85$. Inset: $\ln L^{\beta/\nu_\perp}\rho$ versus $\ln L$ for the same
values of $p$.
\vspace{1em}

\noindent FIG. 2. Quasistationary moment ratio $r_{211;2}$ versus system size
for $p=0.12990$, $p=0.12993$, and $p=0.12995$, from top to bottom $D=0.85$.
\vspace{1em}

\noindent FIG. 3. Moment ratios $ r_{211;2}$ (pairs) and $r_{211;1}$
(particles) versus $\ln L$ for $p=0.120354$ and $D=0.5$. Dashed line: moment
ratio $ r_{211;2}$ for the critical PCP.
\vspace{1em}

\noindent FIG. 4. Scaled histograms for the critical PCPD with $D=0.1$, 0.5,
and 0.85 (from top to bottom).  Inset: scaled histograms for the critical PCPD
($D=0.5$, $L=10240$ and 20480), the critical CP (same sizes) and the critical
PCP ($L=10240$).
\vspace{1em}

\noindent FIG. 5. Quasistationary moment ratio $r_{211;2}$ versus $p$ for
system sizes $L=1280$, $L=2560$...$L=20480$ and $D=0.85$. Inset $\ln{\partial
m/\partial p}$ versus $\ln{L}$ at criticality.
\vspace{1em}

\begin{figure}[h]
\epsfysize=14cm
\epsfxsize=14cm
\centerline{
\epsfbox{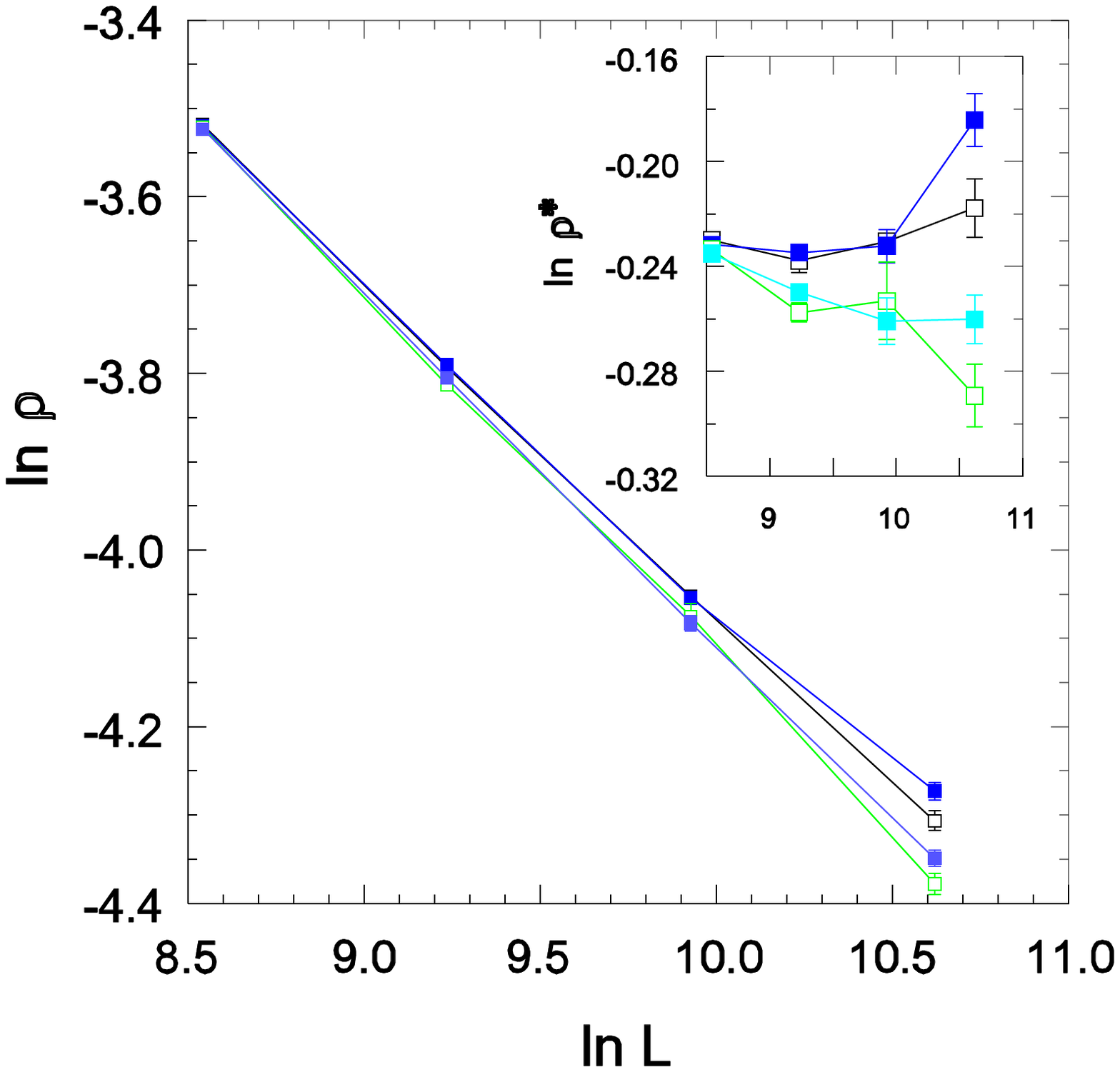}}
\end{figure}

\begin{figure}[h]
\epsfysize=14cm
\epsfxsize=14cm
\centerline{
\epsfbox{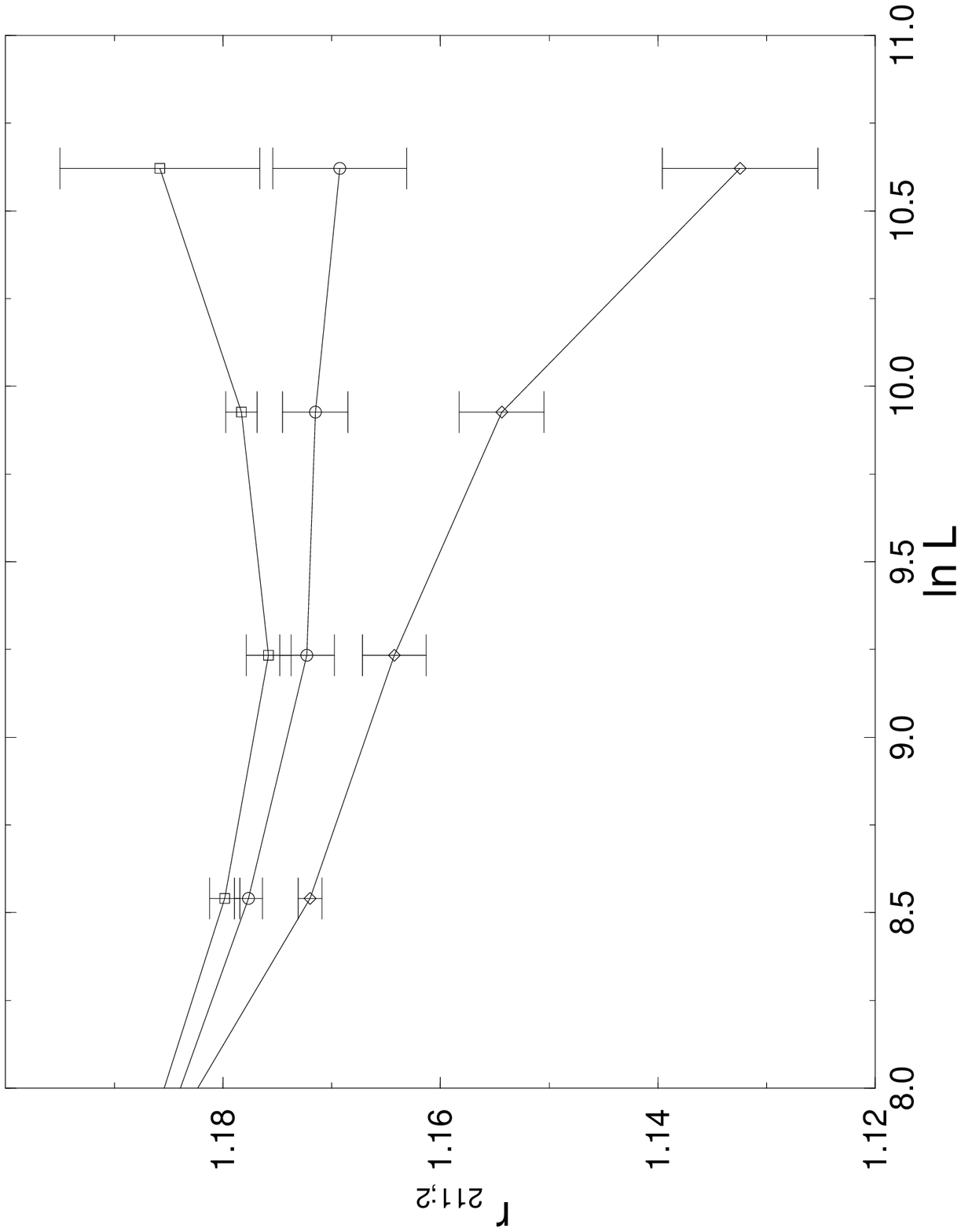}}
\end{figure}

\begin{figure}[h]
\epsfysize=14cm
\epsfxsize=14cm
\centerline{
\epsfbox{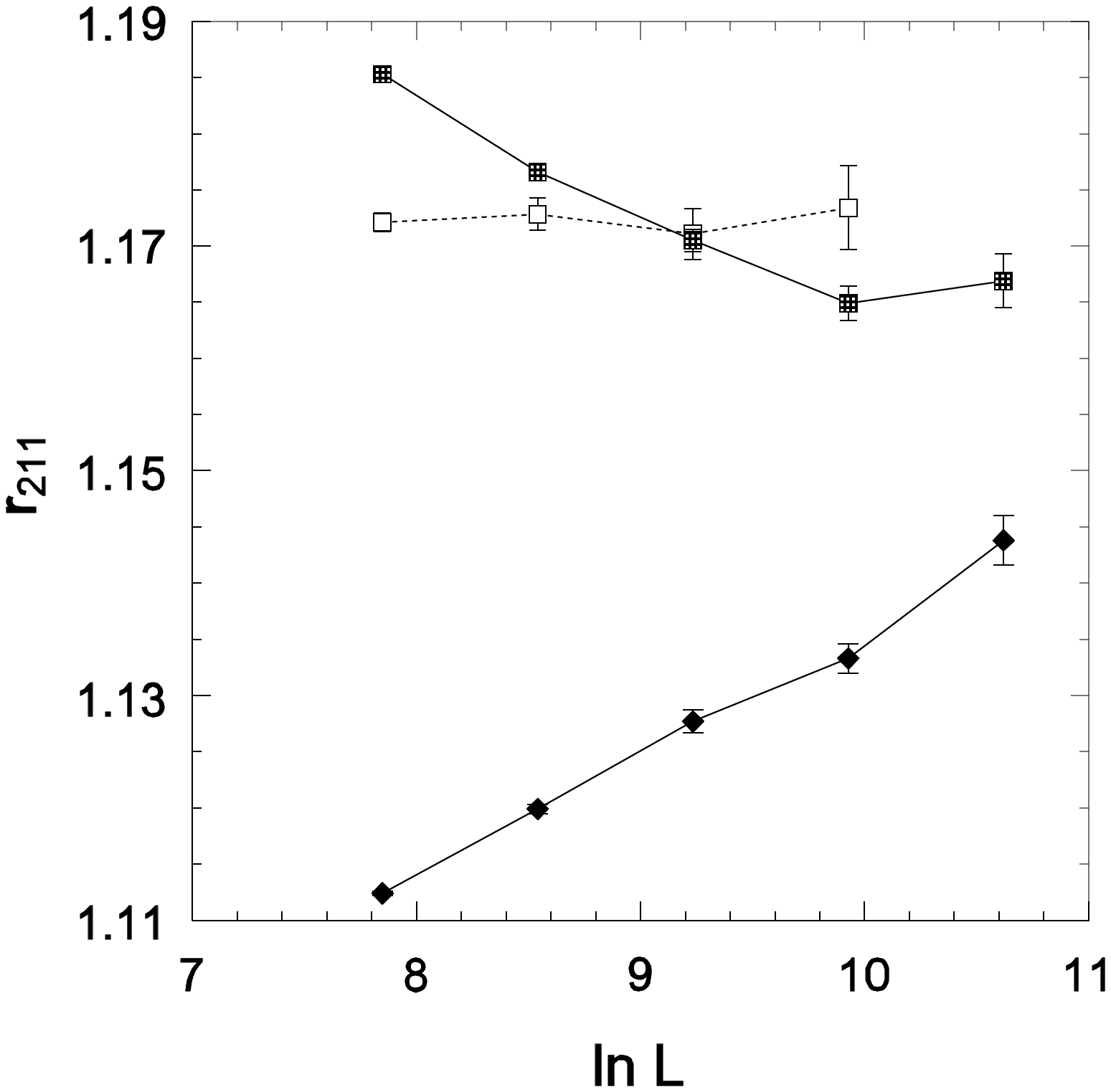}}
\end{figure}

\begin{figure}[h]
\epsfysize=14cm
\epsfxsize=14cm
\centerline{
\epsfbox{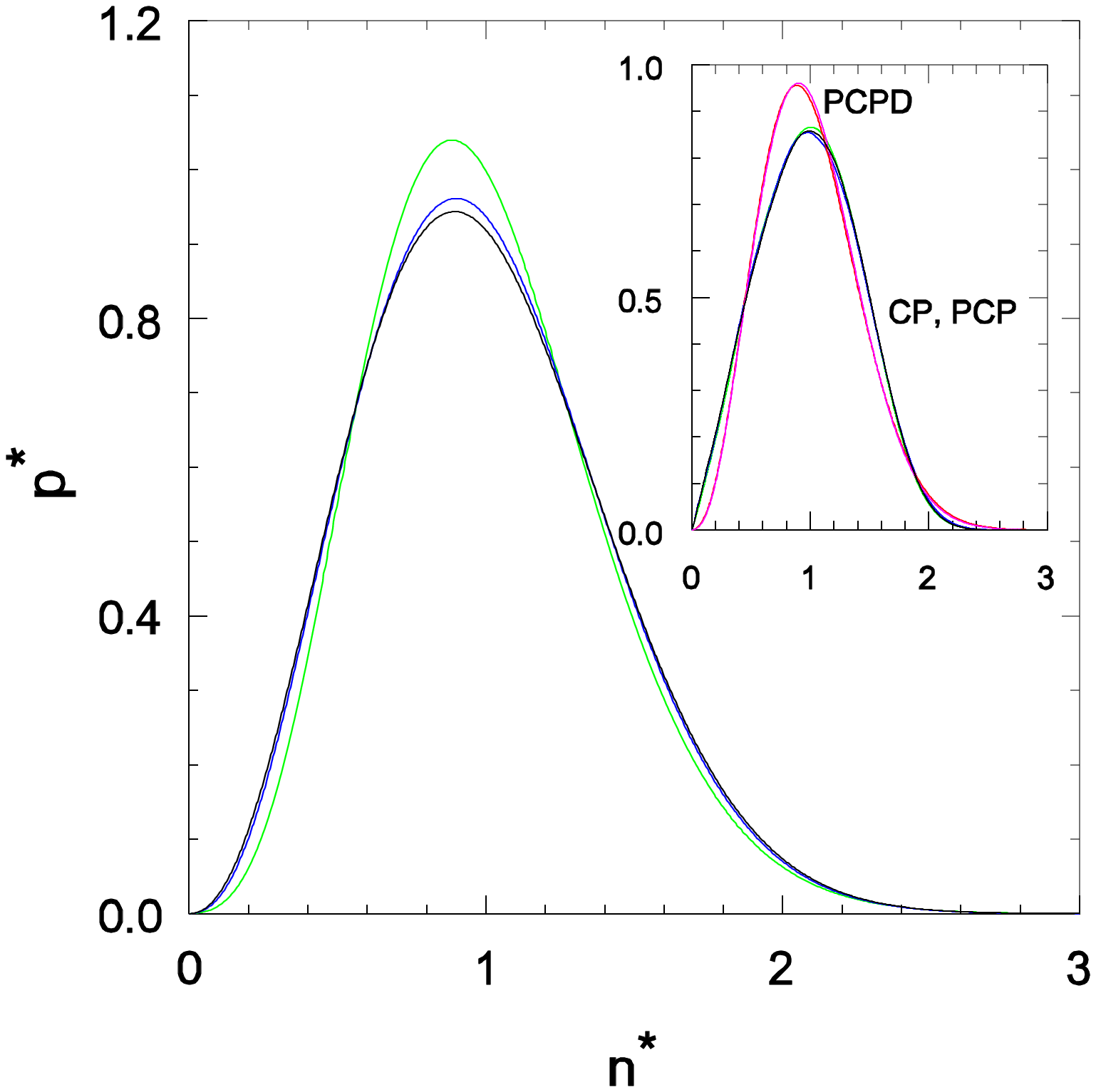}}
\end{figure}

\begin{figure}[h]
\epsfysize=14cm
\epsfxsize=14cm
\centerline{
\epsfbox{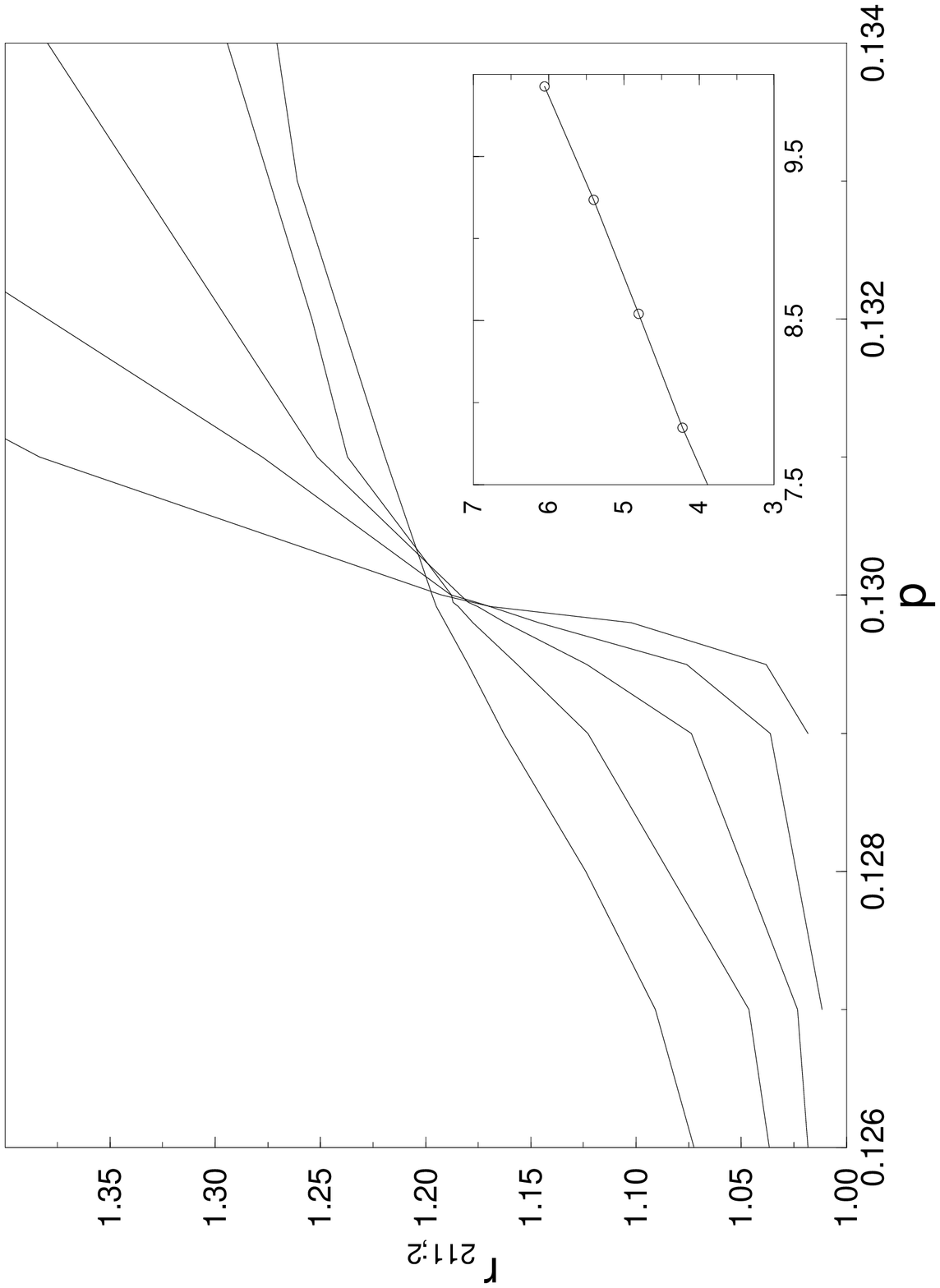}}
\end{figure}

\end{document}